\documentclass[12pt]{revtex4-1}
\usepackage{graphicx}
\usepackage[usenames]{color}
\usepackage{color}
\usepackage[sort&compress]{natbib}

\begin{document}

\title{Fragmentation pathways of nanofractal structures on a surface}
\date{\today}

\author{Veronika V. Dick$^{1}$}
\author{Ilia A. Solov'yov$^{1,2,3}$\footnote{E-mail: ilia@illinois.edu}}
\author{Andrey V. Solov'yov$^{1,2}$}%

\address{$^{1}$Frankfurt Institute for Advanced Studies, Goethe University, Ruth-Moufang-Str. 1, 60438 Frankfurt am Main, Germany}
\address{$^{2}$A.F. Ioffe Physical-Technical Institute, Politechnicheskaya 26, 194021 St. Petersburg, Russia}
\address{$^{3}$Beckman Institute for Advanced Science and Technology, University of Illinois at Urbana-Champaign, 405 N. Matthews Ave., 61801 Urbana, Illinois, USA}

\begin{abstract}
We present a theoretical analysis of the post-growth processes occurring in nanofractals grown on a surface. For this study we have developed a method which accounts for the internal dynamics of particles in a fractal. We demonstrate that the detachment of particles from the fractal and their diffusion within the fractal and over the surface determines the shape of the islands remaining on a surface after the fractal fragmentation. We consider different scenarios of fractal post-growth relaxation and analyze the time evolution of the island's morphology. The results of our calculations are compared with available experimental observations, and experiments in which the post-growth relaxation of deposited nanostructures can be tested are suggested.

\end{abstract}
\maketitle

\section{Introduction}

Nowadays, nanoscience is a rapidly developing research domain \cite{RubahnBook,DrexlerBook,BroerBook, BrechignacBook}. This generic word refers to the study performed on systems having a characteristic length scale on the order of a nanometer: a length scale at which new specific physical and chemical properties emerge in the system. One of the main goals of nanotechnology is the development of controlled, reproducible, and industrially transposable nanostructured materials \cite{RubahnBook,DrexlerBook,BroerBook,BrechignacBook,BarabasiBook,Dugueta2011,Evans2006}. In this context, controlling of the final architecture of such materials by tuneable parameters is one of the fundamental problems.

The conventional technique of thin-film growth by deposition of atoms \cite{BroerBook,DrexlerBook}, small atomic clusters \cite{BrechignacBook,BroerBook} and molecules \cite{RubahnBook,DrexlerBook,VSolovyova09,VSolovyova10a,VSolovyova10b} on surfaces gives a possibility to construct materials with pre-defined properties. Recent experiments show that patterns with different morphology can be formed in the course of cluster deposition on a surface \cite{BrechignacBook,BroerBook,Jensen99}. Among other possible shapes, droplet-like and fractal islands have been observed in various systems \cite{BrechignacBook,BroerBook,Jensen99,Zhang2010,Thiel2009}. It was shown that the island morphology depends on various factors, such as the temperature \cite{BarabasiBook,BrechignacBook,BroerBook, Brechignac06, Brechignac03}, particle size \cite{Brechignac06a}, particle deposition rate \cite{Bardotti95,Scott06,BarabasiBook}, substrate roughness \cite{Brechignac09, Brechignac08}, concentration of impurities in the system \cite{Brechignac06,Brechignac07,BrechignacBook} and interparticle interaction energies \cite{BrechignacBook,BarabasiBook}. It was also demonstrated that the patterns on a surface strongly depend on the type of the substrate. For example, experimental studies of silver clusters deposited on silicon at room temperature showed that droplet-like islands are formed \cite{Bhattacharyya08}, while in \cite{Brechignac03,Brechignac06,Brechignac07} it was demonstrated that dendritic shapes emerge on graphite.

The investigation of the dendritic structures (fractals) has attracted considerable attention \cite{Conti98,Sharon03,Sempere93,Brechignac03,Brechignac06,Brechignac07,Li09,Jensen94,Jensen99}. The formation of such systems provides a natural framework for studying disordered structures on a surface because fractals are generally observed in far-from-equilibrium growth regime. For example, fractals consisting of Ag \cite{Brechignac03,Brechignac06,Brechignac07}, Au \cite{Liu07}, Fe-N \cite{Li09} clusters and C$_{60}$ molecules \cite{Liu06,Kappes05} have been fabricated on different surfaces with the use of the cluster deposition technique \cite{RubahnBook,DrexlerBook}.

The growth process of fractals has been extensively studied in experiments \cite{Bardotti95,Hou98,Liu06,Liu07,Scott06,Brechignac09,Brechignac08}. In \cite{Bardotti95, Scott06} a quantitative experimental study of spherical antimony cluster diffusion on graphite was performed. It was shown that the size of the emerging fractals depends on the cluster deposition rate. The influence of cluster size on fractal morphology was experimentally studied in~\cite{Brechignac06a}. In that work antimony clusters of different size were subsequently deposited on graphite surface, and it was demonstrated that the fractal branch width depends on the size of the deposited clusters. Molecular processes, underlying the C$_{60}$-fractal formation on graphite substrate were investigated experimentally by use of the scanning tunneling microscopy \cite{Liu06}. The self-organization of silver clusters on graphite surfaces with different crystallographic orientations was experimentally investigated in \cite{Brechignac09}. It was shown that the size of the formed fractals depends on the crystallographic planes of graphite, which influences the cluster mobility over a surface.

Contrary to the process of fractal formation, the process of post-growth relaxation and the question of stability of deposited structures are still not well understood. The understanding of the post-growth relaxation processes would allow one to control the self-organization processes of particles on a surface for the purpose of obtaining patterns with predictable morphology. An illustrative example of pattern manipulation was given in~\cite{Hou98} by adding metal impurity to the system. In that work different morphologies of C$_{60}$ films with triangular, dendritic and fractal-like (111)-oriented single-crystal grains were detected by changing the thickness of the pristine fullerene film and the concentration of Ag impurities.

The post-growth transformation of silver cluster fractals to compact droplets on graphite surface was experimentally studied in \cite{Brechignac03,Brechignac06,Brechignac07}. It was demonstrated that depending on the experimental conditions the shape and the size of the stable silver droplets changes significantly \cite{Brechignac07}. In \cite{Brechignac03,Brechignac06,Brechignac07} it was shown that oxidizing of silver clusters results in rapid fragmentation of a fractal, leading to the formation of several compact droplets.

An important characteristic, which determines fractal formation and post-growth relaxation dynamics is the mobility of a cluster on the substrate, which in turn is temperature dependent. Fractals of gold clusters, grown at room temperature on ruthenium substrate undergo a transformation into compact droplets after annealing at 650 K \cite{Hwang91}. Thermal relaxation of silver cluster fractals was experimentally studied in \cite{Brechignac06,Brechignac07}. In these papers it was demonstrated that due to thermal annealing the fractal branch width increases and eventually the fractal breaks into smaller parts.

The dynamics of particles on a surface was also studied theoretically. An efficient theoretical tool for describing particle dynamics on a surface is the diffusion limited aggregation (DLA) method \cite{Witten81}. In this method each particle on a surface moves freely in a random direction until it collides with another particle. In the case of collision both particles stick together and become immobile. The DLA model was used for a qualitative description of the process of fractal formation on a surface \cite{BarabasiBook,Bardotti95,Irisawa95,Brechignac08}.

Two-dimensional theoretical model based on the DLA method has also been developed for the description of thermal relaxation of fractals on surfaces \cite{Irisawa95}. In this model particles are treated as immobile only if they are complectly surrounded by other particles. In all other cases particles are allowed to move along the branches of a fractal with a certain probability \cite{Irisawa95}. The parameter dependent method developed in~\cite{Irisawa95} was used to describe the thermal transformation of a fractal into a droplet. The description of fractal instability in \cite{Irisawa95} was limited only to one particular choice of parameters. However, the correspondence of these parameters to the actual experimental values was not established.

The island size distribution function is a fundamental quantity in the kinetic description of island growth. It has been widely used to characterize the experimentally measured \cite{Stroscio1994,Bardotti1998,Stoldt1999} as well as computed \cite{Beben2001,Amar1994,Royston2009,Bales1994,Thiel2000} surface morphologies. The scaling of the island size distribution, island density, monomer density and other morphology characteristics of the system are often studied as a function of the surface coverage, and as a function of the ratio of the diffusion rate to the deposition rate \cite{Beben2001,Amar1994,Royston2009,Bales1994}. The scaling of the afore mentioned quantities allows to determine important physical parameters describing the kinetic growth processes on a surface, as, e.g., the activation energy and the diffusion constant \cite{Amar1994,Bales1994}. An important characteristic of the island size distribution scaling is the scaling function. The shape of the scaling function is independent of the initial distribution of particles on a surface and is determined by the island's growth mechanism \cite{Kandel1997}. The scaling of the island size distribution emerging during the post-deposition growth processes on a surface was also suggested \cite{Li1997}. The scaling of island morphology characteristics has been performed for various model systems \cite{Beben2001,Amar1994,Royston2009,Bales1994}, and allows to characterize complex kinetic processes occurring on a surface.

In the present paper we make an important step towards understanding of the different evolution scenarios of nanofractals morphology. We present a systematical theoretical analysis of the post-growth processes occurring in nanofractals on a surface. For this study we have developed a method describing the internal dynamics of particles in a fractal with accounting for their diffusion and detachment. We demonstrate that these kinetic processes determine the final shape of the islands on a surface after the post-growth relaxation. We consider different scenarios of fractal relaxation and analyze the time evolution of the island's morphology. The results of our calculations are compared with experimental measurements of the post-growth relaxation of silver cluster fractals on the graphite substrate \cite{Brechignac06,Brechignac03,Brechignac07}. In particular, we analyze the island size distributions calculated at various conditions and different post-growth fragmentation regimes. In conclusion we outline a number of open problems which should be investigated in the future. For instance as a possible next step one could explore the scaling of the island size distributions during the fractal fragmentation processes, as also suggested in \cite{Li1997}. This and many other interesting relevant questions are beyond the scope of the present work and are left for further investigations.

\section{Theoretical methods}

In this section we discuss the theoretical methods used for studying the dynamics of particles on a surface. Computations were performed with the use
of the MBN Explorer computer package, which is developed for structure optimization, simulation of dynamics and growth processes in various nanosystems \cite{ISolovyov08,Semenikhina08,ISolovyov09,JGeng08,AKoshelev03,ISolovyov04,OObolensky05,ISolovyov03,IliaNW09,Dick09,Geng2010}.
Below we describe the general idea of the computational method used in our work and explain how internal dynamics of particles in a fractal has been accounted for.

To study the diffusion of particles over a surface we used a version of the kinetic Monte Carlo (KMC) method \cite{Evans2006,Lu1991,Lo1999,Liu1993,Liu1997}. The KMC method is based on the Monte Carlo algorithm and is widely used for the study of time evolution of various processes occurring in nature \cite{Levi1997,Albano1996,Voter2005,Guo2007,Liu1997,Liu1993,Lo1999}, such as surface diffusion of particles \cite{Meng96}, vacancy diffusion in alloys \cite{Young66}, damage accumulation and amorphization \cite{Pelaz2003} and many others. The processes described using the KMC method always occur with certain predefined rates. Note, that these rates are input in the KMC algorithm, and the method itself cannot predict them. The calculation of the kinetic rates for different processes is usually a nontrivial problem. The kinetic rates are material-dependent parameters of the KMC method, which in the case of particle diffusion over a surface are determined by the atomic composition of particles, substrate material and interparticle interactions. Therefore, by varying values of the kinetic rates, the KMC method can be used to study the dynamical behavior of various molecular systems. Advanced computational methods are often necessary for the kinetic rates calculation. For example, the rate of particle diffusion over a surface can be extracted from molecular dynamics simulation (for a review see, e.g., \cite{Lyalin2009}).

The idea of the KMC method is as follows. The time-evolution of a molecular system is modeled stepwise in time. With a certain probability, at each step of the simulation, the system undergoes a structural transformation. The new configuration of the system is then used as the starting point for the next simulation step. The transformation of the system is governed by the kinetic rates, input into the KMC method. Note, that at each simulation step the system can be transformed into one of several states. Thus, in the KMC method, the probability for the system to attain a certain configuration is proportional to the corresponding kinetic rate. Due to its probabilistic nature, the KMC method allows to study dynamical processes on time scales significantly exceeding the time scales of the conventional molecular dynamics simulation. This method is ideal in the situations when the intermediate details of the dynamical processes are not so essential, and the transition to the final state of the system can be parameterized by a few kinetic rates.

In our studies we used a modification of the conventional KMC method. First, for each particle on a surface we determine the number of possible diffusion directions. Thereby, a particle can either diffuse freely over a surface, or diffuse along the periphery of the already pre-formed structure on a surface. A diffusion direction of the particle is chosen randomly in such a manner that all possible diffusion directions are equally probable. For a given diffusion direction the probability of particle diffusion is then calculated, and the particle is moved in this direction according to the calculated probability. Below we discuss this method in detail.

\subsection{Fractal Growth}
\label{sec:theoryFracGrowth}

\begin{figure}
\includegraphics[width=163mm,clip]{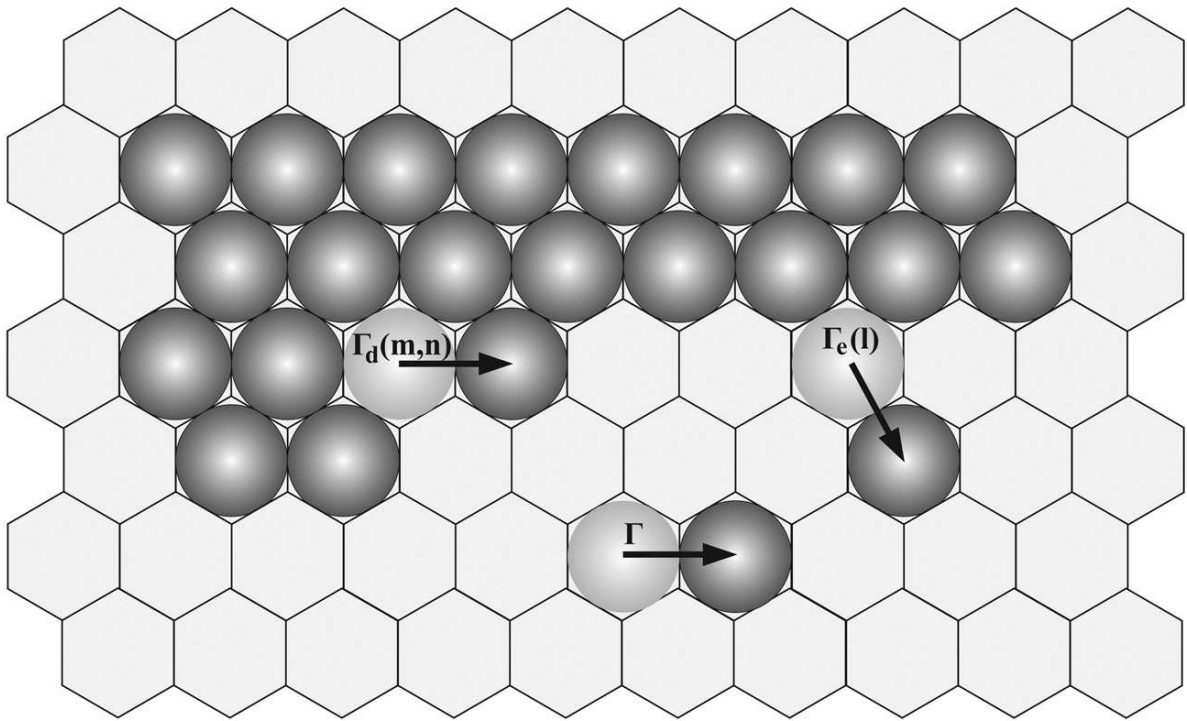}
\caption{Island of deposited particles on hexagonal grid. Processes essential for a fractal formation on a surface are shown by arrows, $\Gamma$ is the free particle diffusion rate, $\Gamma_{d}(m,n)$ is the diffusion rate along the periphery of fractal or island, and $\Gamma_{e}(l)$ is the particle detachment rate from fractal or island. The diffusion rate along the periphery depends on the number of broken bonds $(m)$ and the number of maintained neighboring bonds $(n)$. The particle detachment rate depends on the number of broken bonds $(l)$. In the depicted example $m = 3$, $n = 1$, $l= 2$.}
\label{fig:fig1}	
\end{figure}

To model the growth of a fractal on a surface we used the diffusion limited aggregation (DLA) method \cite{Witten81}. Using a module of the MBN Explorer program \cite{1ISolovyov09} we have computed the growth process of a fractal by depositing particles on a surface in the vicinity of the pre-defined growth center. To compare with the experimental measurements \cite{Brechignac06,Brechignac07,Brechignac03} we have used in our simulations the model parameters consistent with the experiment. Thus, the diameter of a particle has been taken to be $2.5$ nm, which corresponds to the size of an Ag$_{500}$ cluster used in \cite{Brechignac06,Brechignac07,Brechignac03}. The deposition flux has been decreased from $F_{start} = 7.2\times10^{13}$ particle/cm$^{2}$s to $F_{end} = 1.1\times10^{11}$ particle/cm$^{2}$s because the area to which the particles are added decreases as the size of the fractal increases. The used values of the particle deposition flux are chosen higher than the experimental value reported in~\cite{Brechignac06,Brechignac07}, $F = 10^{10}$ particle/cm$^{2}$s, in order to accelerate simulation of the fractal growth. The simulated fractals have been used as initial structure in the investigation of fractal fragmentation. To speed up the calculation, we simulated particle dynamics on a 2D hexagonal grid, on which a particle has up to six neighbors, as illustrated in Fig.~\ref{fig:fig1}. The size of a single grid cell in this case is defined by the particle diameter.

To simulate fractal growth the following procedure has been adopted. At every step of the simulation new particles are deposited on the surface according to the deposition rate and occupy some of the free cells in the grid. Simultaneously, the already deposited particles diffuse over the surface with the rate

\begin{equation}
\Gamma = \nu_{1}\exp\left[-\frac{E_{a}}{kT}\right],
\label{eq:eq1}
\end{equation}

\noindent
where $E_{a}$ is the activation energy, $\nu_{1}$ is the attempt escape rate, $T$ is the temperature of the system and $k$ is the Boltzmann constant. The process of particle diffusion over the surface is schematically illustrated in Fig.~\ref{fig:fig1}.

An important quantity in the DLA method is the time step, $\Delta t$, which defines the characteristic time for particle diffusion over a surface as

\begin{equation}
\Delta t = 1/\Gamma.
\label{eq:DeltaT}
\end{equation}

\noindent
The time step $\Delta t$ is related to the coefficient $D$ of particle's diffusion over a surface arising in the equation of diffusion \cite{EinsteinBook,LL6}. The solution of the diffusion equation in two dimensions gives the probability to find a particle at the instant $t$ being at distance [$r$, $r+dr$] from its initial position as

\begin{equation}
\omega(r,t)dr=\frac{1}{2Dt}\exp\left(-\frac{r^2}{4Dt}\right)rdr.
\label{eq:probFunc}
\end{equation}

\noindent
Using this probability function one derives the mean square displacement of a particle as \cite{EinsteinBook,LL6}

\begin{equation}
\overline{(r_1-r_0)^2}=\int_0^{\infty}\omega(r,t_1-t_0)r^2dr=4D(t_1-t_0),
\label{eq:sqrDisp}
\end{equation}

\noindent
where $r_0$ and $r_1$ are the distances to a particle from the initial position at two successive instances $t_0$ and $t_1$. Equation~(\ref{eq:sqrDisp}) allows to express the diffusion coefficient as

\begin{equation}
D = \frac{\langle\Delta r^{2}\rangle}{2z \Delta t},
\label{eq:eq2}
\end{equation}

\noindent
where $\langle\Delta r^{2}\rangle$ is the mean-square displacement of a particle per time $\Delta t$, and $z$ is defined by the dimensionality of space \cite{EinsteinBook,LL6}. In the case of particle diffusion over a surface $z=2$ (see Eq.~(\ref{eq:sqrDisp})).

On the other hand the mean-square displacement depends on the diffusion rate and on the particle hopping length, which is in the example considered is equal to the particle diameter $d_{0}$:

\begin{equation}
\langle\Delta r^{2}\rangle = \Gamma d^{2}_{0} \Delta t.
\label{eq:eq3}
\end{equation}

\noindent
Here $\Delta t$ has a meaning of a single simulation step defined in Eq.~(\ref{eq:DeltaT}). Substituting Eq.~(\ref{eq:eq3}) into Eq.~(\ref{eq:eq2}), one derives

\begin{equation}
D = \frac{\Gamma d^{2}_{0}}{2z}.
\label{eq:eq4}
\end{equation}

\noindent
Equation~(\ref{eq:eq4}) allows to estimate $\Gamma$ (and therefore $\Delta t$) once the diffusion coefficient is known:

\begin{equation}
\Delta t = \frac{d^{2}_{0}}{2z D}.
\label{eq:eq5}
\end{equation}

The diffusion coefficient of an Ag$_{500}$ cluster on graphite at room temperature was measured as $2\cdot10^{-7}$~cm$^{2}$/s \cite{Brechignac07}. Substituting this value into Eq.~(\ref{eq:eq5}), one obtains $\Delta t = 78$~ns.

Substituting Eq.~(\ref{eq:eq1}) into Eq.~(\ref{eq:eq4}), one relates the diffusion coefficient to the activation energy and temperature:

\begin{equation}
D = \frac{d^{2}_{0} \nu_{1}}{2z} \exp\left[-\frac{E_{a}}{kT}\right].
\label{eq:eq6}
\end{equation}

\noindent
It follows from Eq.~(\ref{eq:eq6}) that the diffusion coefficient decreases as the activation energy grows. This results in an exponential growth of the time step $\Delta t$ with $E_a$, since $\Delta t\sim1/D$ (see Eq.~(\ref{eq:eq5})). Equation~(\ref{eq:eq5}) introduces the optimal time step for the computations, because it defines the characteristic time at which a freely deposited particle gets displaced for the distance $d_0$, i.e. to the neighboring lattice cell (see Fig.~\ref{fig:fig1}). Note, that the diffusion of deposited particles is typically the fastest process in the system. Therefore, it determines the minimum time step and the time scale for the growth and fragmentation of fractal structures on a surface. It is computationally inefficient to perform simulation with time steps much less than $\Delta t$, because in this case the particles will practically not move during $t\ll\Delta t$.

In our method we have employed the following procedure to model particle dynamics over a surface: for each step of the simulation the deposited particles without neighbors have six possibilities for diffusion, see Fig.~\ref{fig:fig1}. The direction of displacement is chosen randomly and each particle is moved to a neighboring lattice cell in the chosen direction. Thereby each step of the simulation corresponds to $\Delta t=78$~ns, as estimated above. The particles at the fractal periphery diffuse slower, as shown in the next section, and therefore are displaced less frequently.

\subsection{Kinetic processes in fractal fragmentation}
\label{sec:kinProc}

In this work we consider fragmentation of a two-dimensional fractal consisting of rigid particles of the equal radius. The relaxation of such fractal is driven by the diffusion of particles along the fractal periphery and by the detachment of particles from the fractal. Both processes are schematically depicted in Fig.~\ref{fig:fig1}. The diffusion and the detachment rates depend on the activation energy and the particle-particle interaction. In the Arrhenius approximation the diffusion rate of a particle along the fractal periphery reads as:

\begin{equation}
\Gamma_{d} (m,n) = \nu_{2} \exp\left[-\frac{mE_{b}}{kT} -\frac{n \Delta \epsilon}{kT} - \frac{E_{a}}{kT}\right],
\label{eq:eq7}
\end{equation}

\noindent
where $m$ is the number of bonds that are broken due to the particle motion, $E_{b}>0$  is the binding energy between two particles, $n$ is the number of maintained neighboring bonds between two particles and $\Delta\epsilon \leq E_{b}$ is the diffusion energy barrier \cite{Irisawa95,Brechignac03}, $\nu_{2}$ is the attempt escape rate. Equation~(\ref{eq:eq7}) describes the probability of a particle to overcome a potential energy barrier, which for a particle diffusing along the fractal periphery is parameterized by the energies $E_{b}$, $\Delta\epsilon$, and $E_a$. Note that the parameter $E_a$, which enters Eq.~(\ref{eq:eq7}), depends on the simulation time step $\Delta t$, as discussed in the previous section. Therefore, only the parameters $E_b$ and $\Delta\epsilon$ define the potential energy barrier for particle diffusion along the fractal periphery, while $E_a$ characterizes the time scale.

Note that Eq.~(\ref{eq:eq7}) does not account for the bonds which may be created in the system when a particle diffuses. This feature of Eq.~(\ref{eq:eq7}) is easy to understand. Indeed, the particle diffusion process in our model is considered stepwise, i.e., at each step of the computation a particle is displaced with a certain probability in a random direction for the distance equal to its diameter. But prior the particle is displaced to its new position there is no information about the newly created bonds in the system (causality principle). Therefore only those bonds which the particle forms with its neighbors prior the
displacement influence the diffusion dynamics in the system.

The evaporation (detachment) rate of a particle from the fractal is given by

\begin{equation}
\Gamma_{e} (l) = \nu_{3}\exp\left[-\frac{lE_{b}}{kT} -\frac{\Delta\mu}{kT} - \frac{E_{a}}{kT}\right],
\label{eq:eq8}
\end{equation}

\noindent
where $l$ is the number of bonds broken during the particle detachment from fractal, $\Delta\mu$ is the chemical potential
change associated with particle detachment~\cite{Irisawa95,Brechignac03,BrechignacBook,FrenkelBook,Metiu1997}, $\nu_{3}$ is the attempt escape rate of a
particle in its equilibrium state. Equation~(\ref{eq:eq8}) can be understood within the framework of the classical nucleation theory \cite{FrenkelBook}, which studies the liquid$\leftrightarrow$gas transition in droplets. It is written in the Arrhenius approximation, similarly to Eq.~(\ref{eq:eq7}). For the further discussion of the fractal fragmentation we put:

\begin{equation}
\nu_{2}\simeq\nu_{3} = \nu.
\label{eq:eq9}
\end{equation}

\noindent
Such situation occurs when the characteristic attempt escape rate of a particle leading to its diffusion or detachment are close. Equations~(\ref{eq:eq7})-(\ref{eq:eq8}) describe the dependence of the probability of different essential kinetic processes on the values of $E_{a}$, $E_{b}$, $\Delta\epsilon$, $\Delta\mu$, which below are called the model parameters. For convenience, in this paper we define all the model parameters in units of $kT$ ($1~kT = 0.026$~eV) at room temperature (300~K).

\section{Results}


\subsection{Model parameters}
\label{sec:resKinParam}

The interaction energy between the deposited particles and the substrate is responsible for the particle mobility over a surface, as follows from Eq.~(\ref{eq:eq1}). {The energetic parameters for the atomic-scale processes on the Ag(100) surface were studied in \cite{Thiel2004}, while the energetic parameters for Si atom migration on Si(100)-2$\times$1 surface were discussed in \cite{Lu1991}.} The interaction energy of Ag$_{500}$ ($E^{Ag}_{a}$), C$_{60}$ ($E^{C_{60}}_{a}$), and Sb$_{2300}$, ($E^{Sb}_{a}$) clusters with graphite surface at room temperature has been estimated as $E^{Ag}_{a} = 6.6$~$kT$ \cite{Kebaili09}, $E^{C_{60}}_{a} = 6.9$~$kT$ \cite{Liu08} and $E^{Sb}_{a} = 27.1$~$kT$ \cite{Bardotti95}. The significant spread of the values indicates the essential role of interatomic interactions in defining the activation energy. The value of $E_a$ defines the time scale of the fractal growth and fragmentation processes, as discussed in section~\ref{sec:theoryFracGrowth}. In this paper we describe the dynamics of silver cluster fractals with the above mentioned value of $E_a$.
Another important quantity characterizing the particle diffusion over a surface is its attempt escape rate $\nu$ (see Eqs.~(\ref{eq:eq7})-(\ref{eq:eq9})), which can be estimated as

\begin{equation}
\nu = \frac{2Dz}{d^{2}_{0}}\exp\left[\frac{E_{a}}{kT}\right].
\label{eq:eq10}
\end{equation}

\noindent
For a silver nanoparticle with $d_0=2.5$~nm deposited on graphite the diffusion coefficient at room temperature $D\simeq2\cdot10^{-7}$~cm$^{2}$s$^{-1}$ \cite{Brechignac07}, resulting in $\nu=0.94\cdot10^{10}$~s$^{-1}$. {Note that the attempt escape rate for silver clusters is 2-3 orders of magnitude smaller than for individual atoms on a surface \cite{Evans2006,Lu1991,Lo1999,Thiel2004}.}

The interaction energy between two particles, $E_{b}$, depends on the atomic composition of the particles and on the presence of impurities in the system \cite{Brechignac06,Brechignac07,Brechignac03}. Thus, it was shown that the presence of oxygen impurities in a silver cluster deposited on graphite leads to the decrease of $E_{b}$ and consequently to the reduction of fractal stability. A systematic study of the afore mentioned factors on the interparticle interaction energy is beyond the scope of this paper and deserves a separate systematic investigation. Here for us is important that according to experiment \cite{Brechignac06,Brechignac07} silver cluster fractals are formed and decay on the comparable time scales. This means that in our model $E_{b}$ should be of the same order of magnitude as $E_{a}$. It is worth noting that the time of fractal formation can become significantly smaller than the time of fractal fragmentation if the conditions at which the fractal is kept after growth rapidly change, e.g. the temperature of the system is increased. The diffusion barrier energy $\Delta \epsilon$ depends on the atomic composition of the cluster and usually amounts $0.05-0.2$ of the bonding energy of two clusters \cite{OuraBook}.

The change in the chemical potential $\Delta \mu$ arises due to the energy difference caused by the change of the number of particles in the system. The chemical potential characterizes the ability of particles to diffuse from regions of high chemical potential to those of low chemical potential and is defined as the partial derivative \cite{LL5}

\begin{equation}
\mu=\left(\frac{\partial U}{\partial N}\right)_{V,S},
\label{eq:chemPot}
\end{equation}

\noindent
where $U$ and $S$ are the total energy and the entropy of the system, $V$ is its volume and $N$ is the number of particles in the system. The variation of the chemical potential arising due to a structural transformation in the system can be calculated from the known values of the chemical potential of individual components of the system before and after the transformation. For example, for the evaporation of a silver nanoparticle from a fractal with $N$ particles on graphite surface

\begin{equation}
{\rm Ag(fractal_{N})}+{\rm C(graphite)}\rightarrow{\rm Ag(fractal_{N-1})}+{\rm C(graphite)}+{\rm Ag(particle)},
\label{eq:reaction}
\end{equation}

\noindent
the corresponding change of the chemical potential can be calculated as a difference between the chemical potential of the products and the educts. With $\mu_{\rm Ag(fractal_{N})}\approx\mu_{\rm Ag(fractal_{N-1})}$ one obtains

\begin{equation}
\Delta\mu=\mu_{\rm Ag(particle)}.
\label{eq:deltaMu}
\end{equation}

\noindent
The chemical potential can be measured experimentally \cite{Job06} and is tabulated for many substances (see e.g. \cite{Wiberg72,Stull71}). It depends on the phase state of the system: for the gas of silver atoms $\mu_{{\rm Ag}}^{{\rm (gas)}}=2.55$~eV, while for the silver in the liquid phase $\mu_{{\rm Ag}}^{{\rm (liquid)}}=0.8$~eV \cite{Wiberg72}. These values and Eq.~(\ref{eq:deltaMu}) allow one to suggest that the change of the chemical potential in the silver fractal fragmentation process, at room temperature lies within the range $30-100$~$kT$.

\subsection{Fractal growth}

{We stress that this paper is devoted to the process of fractal post-growth evolution. The problem of nanofractal formation has been studied in many papers (see e.g. \cite{BarabasiBook,Bardotti95,Irisawa95,Brechignac08,Hou98,Liu06,Liu07,Scott06,Brechignac09}) and we intend to extend this analysis in a separate publication. Therefore, here we give only a hint how the fractals of the morphology of interest can be obtained.}

\begin{figure}[htb]
\includegraphics[width=163mm,clip]{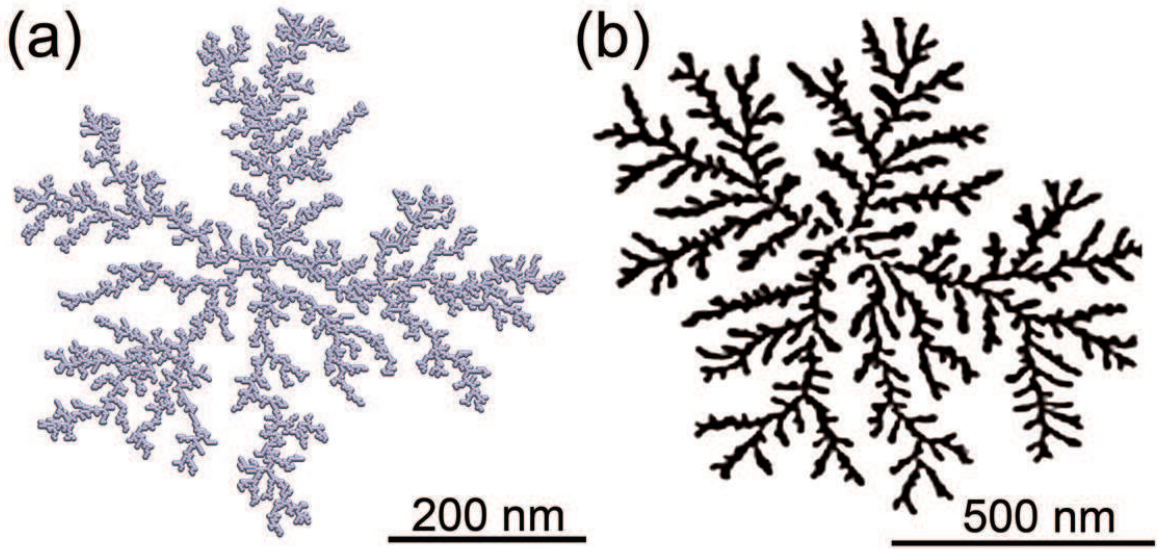}
\caption{(a) Fractal structure simulated using the DLA method; (b) structure of silver cluster fractal experimentally grown by clusters deposition technique on graphite surface~\cite{Brechignac03};}
\label{fig:fig2}
\end{figure}

Using the method described in Sec.~IIA we have simulated several fractals with the structure being very similar to the silver cluster fractals grown experimentally on graphite surface \cite{Brechignac06,Brechignac03,Brechignac07}. The fractal structure shown in Fig.~\ref{fig:fig2}a have been chosen for the further investigation of the post-growth relaxation processes in fractals. The diameter of the fractal is $635$~nm, which is somewhat smaller than the diameter of the experimentally grown structures~\cite{Brechignac06,Brechignac03,Brechignac07}. For the sake of illustration in Fig.~\ref{fig:fig2}b we show the experimentally grown silver cluster fractal prior thermal annealing, which triggers fractal fragmentation~\cite{Brechignac06,Brechignac03,Brechignac07}.

The important characteristic of a fractal is the fractal dimension $d_{f}$. The Hausdorff fractal dimension is generally defined as \cite{Hausdorff1919,Martinez01}:

\begin{equation}
d_{f} = \lim_{l \rightarrow 0} \frac{\log[N(l)]}{\log[1/l]}.
\label{eq:eq15}
\end{equation}

\noindent
Here $N(l)$ is the number of self-similar structures of linear size $l$ needed to cover the whole structure. In practice the fractal dimension is usually calculated by the box-counting method \cite{Feder}. Equation~(\ref{eq:eq15}) has been used to calculate the fractal dimension of the structure shown in Fig.~\ref{fig:fig2}a. This calculation resulted in $d^{th}_{f} = 1.76$. This value is in a good agrement with experiment for silver cluster fractals grown on the graphite surface, which gives $d^{exp}_{f} = 1.7 \pm 0.1$ \cite{Brechignac06}.

As illustrated in Fig.~\ref{fig:fig2} the topology of fractals simulated by the DLA method is very similar to the fractal topology grown in experiment. In both cases the fractals shown in Fig.~\ref{fig:fig2} have several main branches, growing from the center of the fractal. The branch width of the fractal simulated by the DLA method is $\sim10$~nm, while the typical experimental width of the branch is $15-30$~nm \cite{Brechignac06}. The difference arises because the particles in the simulation were deposited on a surface at a higher rate than those in experiment (see section~\ref{sec:theoryFracGrowth} for details). In addition, the implemented DLA method does not allow the deposited particles to be placed atop of a growing fractal. Another factor affecting this difference is the sticking probability of the deposited particles assumed to be equal to one, meaning that if a particle meets another particle on a surface the two particles stick and do not move together. This is probably not the case in experiment, where the sticking probability can be lower than one and the mobility of complexes with two or more particles is no equal to zero. The sticking probability is less important in the fractal fragmentation process as this process is mainly governed by the detachment rate $\Gamma_e$, and the diffusion rate $\Gamma_d$ introduced in section~\ref{sec:kinProc}. Since the main focus of this paper is the investigation of pathways of fractal fragmentation, we do not discuss further the effect of sticking probability on pattern formation.

\subsection{Fractals fragmentation}

In this section we perform analysis of the fractal post-growth relaxation using the method described in Sec.~\ref{sec:kinProc}. According to the estimates performed in Sec.~\ref{sec:theoryFracGrowth}, a single time step in our calculation is equal to $\Delta t = 78$~ns, which allows one to simulate the process during the time period

\begin{figure}[t]
\includegraphics[width=163mm,clip]{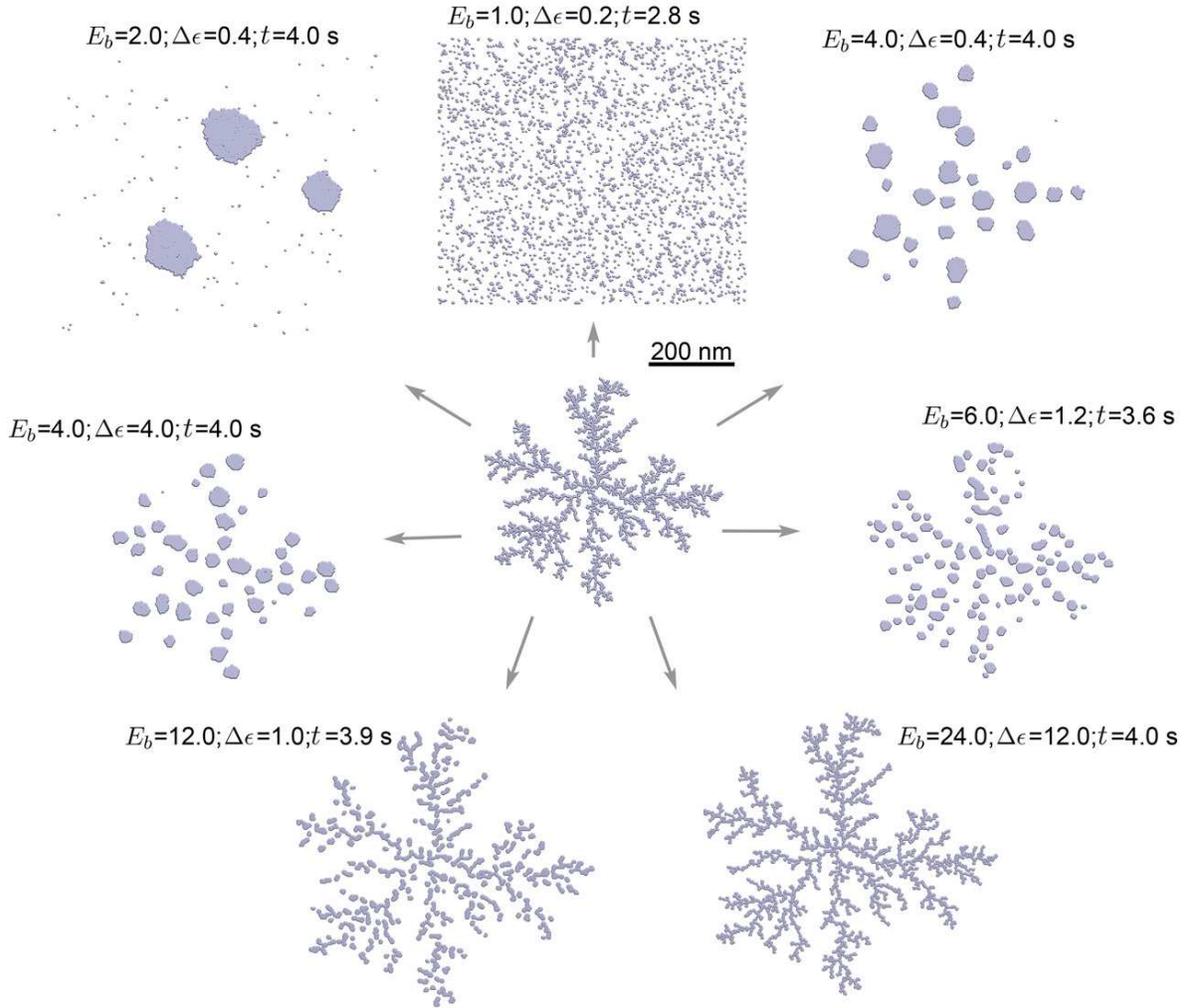}
\caption{Evolution of fractal structure on a $650\times750$~nm$^2$ substrate with periodic boundary conditions. The initial fractal structure shown in the middle undergoes fragmentation in different final states depending on the interparticle interactions in the system. Numbers above the corresponding images indicate the values of $E_b$ and $\Delta\epsilon$ used in the simulations (in units of $kT$), $\Delta\mu=2$~$kT$ in all cases. The simulation time $t$ is given for each path of the fragmentation.}
\label{fig:fig3}
\end{figure}

\begin{equation}
t = N_{step}\Delta t,
\label{eq:simTime}
\end{equation}

\noindent
where $N_{step}$ is the number of simulation steps.

In the present work we analyze several paths of fractal fragmentation. The rate of fractal decay depends on the interparticle interaction, and it defines the morphology of the fragments that are formed during the process. Snapshots of the structures arising at different stages of the fragmentation process simulated at different parameters of interparticle interactions are shown in Fig.~\ref{fig:fig3}. This example shows how different can be the fragmentation paths and the fragments morphology.

Figure~\ref{fig:fig3} shows that for $E_b=1$~$kT$, $\Delta\epsilon=0.2$~$kT$ one observes an entire defragmentation of a fractal, which is the fastest fragmentation path. In this case the interaction energy between the particles is relatively weak and the probability to evaporate a particle from the fractal is much higher than the probability of newly deposited particles to nucleate. This fragmentation scenario can be realized in experiment if the temperature of the system is rapidly elevated after the fractal was created.

Figure~\ref{fig:fig3} shows that for $E_b\ge2$~$kT$ the fractal melts in a number of compact droplets. Depending on the energies of interparticle interactions the shape of the droplets becomes different. Thus, for $E_b=2$~$kT$, $\Delta\epsilon=0.4$~$kT$ three large, almost spherical, droplets of a similar size are formed. In this case the binding energy $E_b$ between the particles is rather small, allowing relatively easy detachment of particles, but at the same time it is large enough to make the characteristic particle detachment time comparable with the characteristic particle nucleation time, thereby preventing the system from entire defragmentation, observed at $E_b=1$~$kT$. Thus, the fragmentation path at $E_b=2$~$kT$ goes via the rearrangement of the entire system, and the formation of large stable droplets.

A further increase of the interparticle interaction energy leads to the change of the fractal fragmentation pattern. As seen in Fig.~\ref{fig:fig3} at $E_b=4-6$~$kT$ the fractal fragments into several compact droplets. The analysis of morphology of the created patterns leads us to the following main conclusions: (i) the growth of $E_b$ leads to the increase of the number of droplets on a surface (see $E_b=4$~$kT$ and $E_b=6$~$kT$) and to the decrease of their average size. This happens because the detachment of particles from the fractal becomes energetically an unfavorable process, and the fractal fragments mainly due to the peripheral diffusion of particles, initiated at the peripheral defect sites. (ii) The increase of the peripheral diffusion barrier energy $\Delta\epsilon$ suppresses the diffusion of particles, resulting in a slower evolution and fragmentation of the fractal shape. It is remarkable that at $E_b=6$~$kT$ and $\Delta\epsilon=1.2$~$kT$ one observes the formation of elongated islands on a surface which follow the direction of the fractal branches. A further increase of the interparticle binding energy with the simultaneous lowering the barrier energy for the  particle peripheral diffusion favors the formation of elongated islands on a surface. Figure~\ref{fig:fig3} illustrates this for $E_b=12$~$kT$ and $\Delta\epsilon=1$~$kT$. In this case the timescale for the particles to detach from the fractal is significantly larger than that for the peripheral particle diffusion.

A simultaneous increase of the interparticle binding energy and the barrier energy  for the particle peripheral diffusion leads to the growth of the fractal life time. Figure~\ref{fig:fig3} shows that for $E_b=24$~$kT$ and $\Delta\epsilon=12$~$kT$ the fractal has no noticeable changes in its morphology after 4~s of simulation. In the case when the interparticle energies are large, the fractal fragmentation is expected to occur on a larger time scale and can be simulated numerically if the value of the simulation time step is increased.

\begin{figure}
\includegraphics[width=163mm,clip]{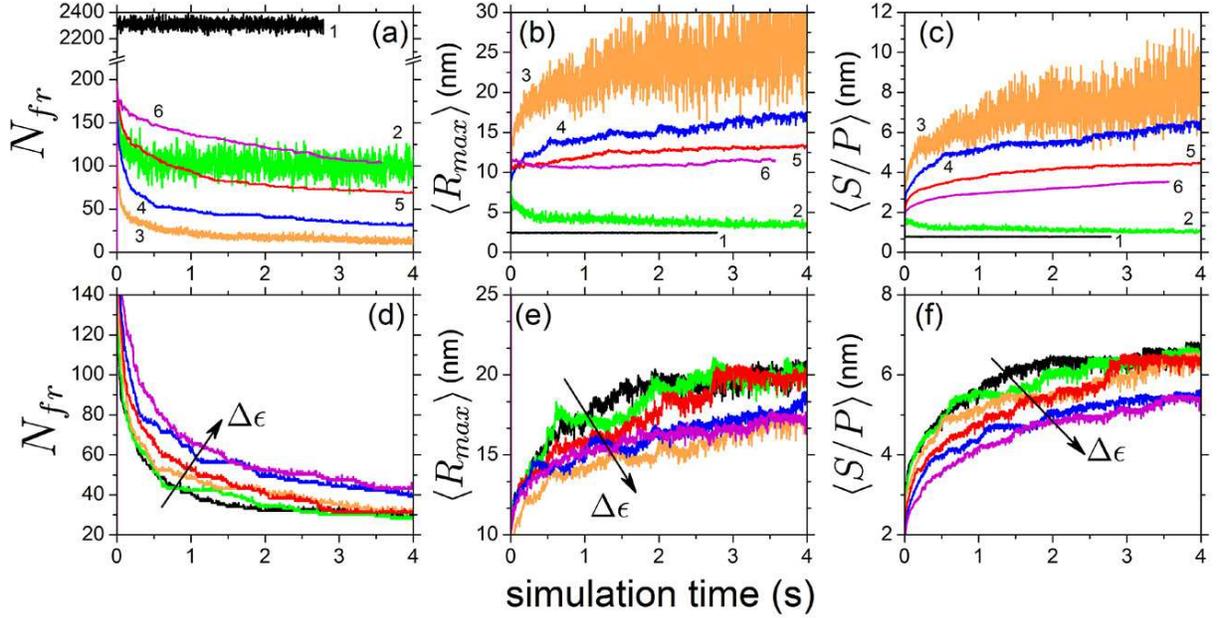}
\caption{Time evolution of the number of fragments $N_{fr}$, $\langle R_{max}\rangle$ introduced in Eq.~(\ref{eq:Rmax}) and of the $\langle S/P \rangle$ ratio introduced in Eq.~(\ref{eq:eq21}) calculated for the fractal structure shown in Fig.~\ref{fig:fig3}. The fractal fragmentation have been analyzed at $\Delta\mu = 2$ $kT$ for different values of the binding energy $E_{b}$ and the barrier energy $\Delta\epsilon$. Plots $(a)-(c)$ show the results of calculation obtained for $\Delta\epsilon=0.2E_{b}$ and the different values of the binding energies between two particles. Lines 1-6 correspond to $E_{b} = (1,~2,~3,~4,~5,~6)~ kT$, respectively. Plots $(d)-(f)$ represent the results obtained at $E_{b} = 4~kT$ for different values $\Delta\epsilon = (0,~0.4,~0.8,~1.0,~3.2,~4)$~$kT$. The direction of growth of $\Delta\epsilon$ is shown in these plots.}
\label{fig:fig4}
\end{figure}

The important characteristic of the fractal fragmentation is the number of fragments at a given time. The smallest fragment is a single particle. The time evolution of the number of fragments calculated for different sets of model parameters is shown in Fig.~\ref{fig:fig4}a and Fig.~\ref{fig:fig4}d. Curve 1 in Fig.~\ref{fig:fig4}a shows the time evolution of the number of fractal fragments at $E_b=1$~$kT$. The number of fragments in this case rapidly approaches the asymptotic value, approximately equal to the half of the total number of particles in the fractal. This means that the system dominantly consists of dimers. With increasing $E_b$ the number of fragments at the equilibrium decreases, as seen in Fig.~\ref{fig:fig3}. It is interesting to note that at $E_b=2$~$kT$ there are three dominating large islands (see Fig.~\ref{fig:fig3}). The total number of fragments at the end of the simulation in this case is equal to 100, being much smaller than the total number of particles in the system. This feature arises in the situation when a large number of single particles detach from the large droplets but later stick back. In this case the number of single particles fluctuates rapidly resulting in the large fluctuations of $N_{fr}(t)$ dependence shown in Fig.~\ref{fig:fig4}a by curve 2. These results have been calculated for a fractal on a $650\times750$~nm$^2$ substrate with periodic boundary conditions.

Figure~\ref{fig:fig4}d shows that there is no dramatic change in $N_{fr}(t)$ dependence with the growth of $\Delta\epsilon$ at a constant value $E_b$. This analysis shows also that the growth of $\Delta\epsilon$ preventing particles peripheral diffusion hinders the fast transformation of droplets into compact islands which eventually results in the increase of the number of fragments on a surface.

\begin{figure}[t]
\includegraphics[width=163mm,clip]{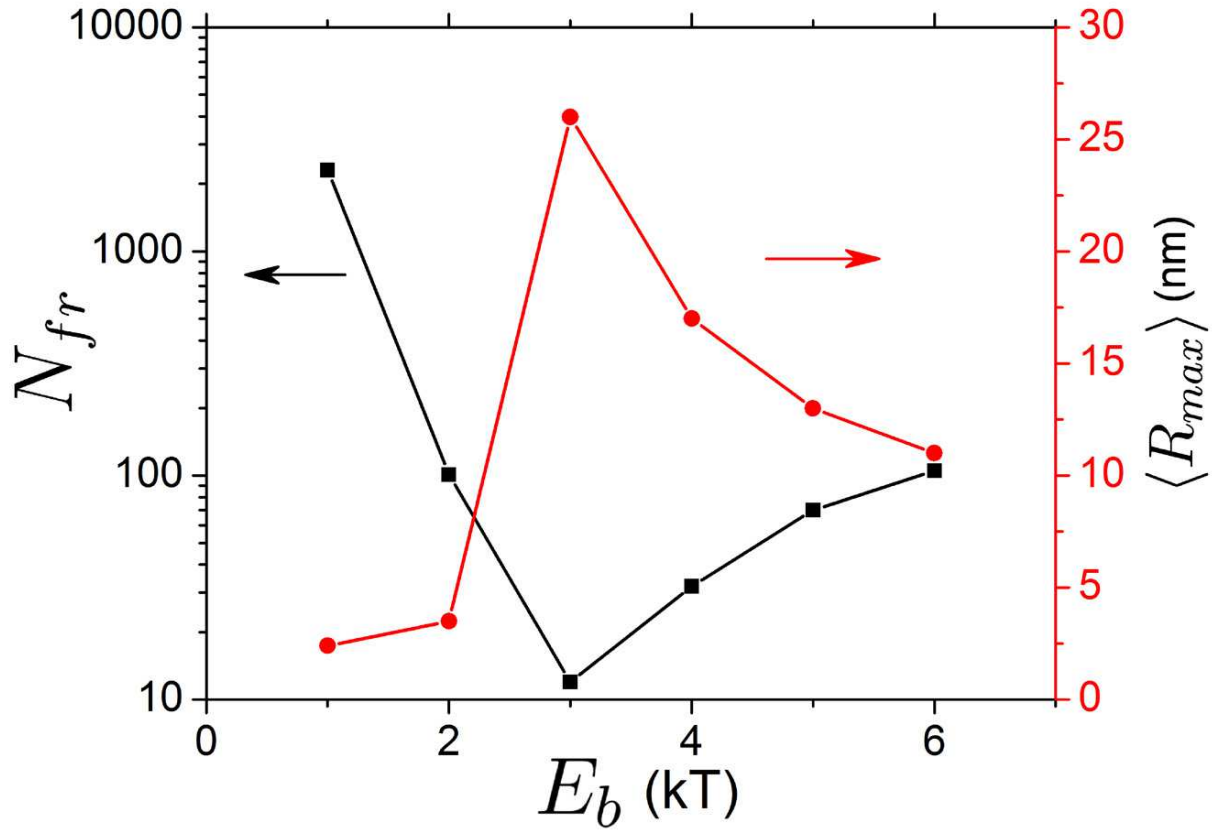}
\caption{{Dependence of $N_{fr}$ (squares, left scale) and $\langle R_{max}\rangle$ (dots, right scale) on the binding energy $E_b$ calculated for the barrier energy $\Delta\epsilon=0.2E_b$ after 4~s simulation, corresponding to the dependencies shown in Fig.~\ref{fig:fig4}a and Fig.~\ref{fig:fig4}b.}}
\label{fig:fig4a}
\end{figure}

As seen in Fig.~\ref{fig:fig3}, in the course of fractal fragmentation the mobile particles can coalescence into islands, i.e. groups of particles bound together. The size and the number of islands on the substrate depend on the binding energy $E_b$ and the barrier energy $\Delta\epsilon$. The important characteristic of the fragmentation pattern on a surface is the average maximal radius of the created islands which reads as

\begin{equation}
\langle R_{max}\rangle = \frac{1}{N_{fr}} \sum^{N_{fr}}_{i=1} R_{max}^{(i)},
\label{eq:Rmax}
\end{equation}

\noindent
where $N_{fr}$ is the total number of islands on a surface, $R_{max}^{(i)}$ is the maximal radius of the $i$-th island. The dependencies of $\langle R_{max}\rangle(t)$ calculated at different values of $E_b$ and $\Delta\epsilon$ are shown in Fig.~\ref{fig:fig4}b and Fig.~\ref{fig:fig4}e. These figures show that in average $\langle R_{max}\rangle$ approaches the equilibrium value at the chosen values of model parameters except for $E_b=3$~$kT$, $\Delta\epsilon=0.6$~$kT$ when the large fluctuations of $\langle R_{max}\rangle$ develop and grow with time. This happens because at $E_b=3$~$kT$ the rate of single particle detachment turns out to be so that only several particles are able to overcome the detachment energy barrier at one simulation step. The escaped particles freely diffusing over the surface after a short period of time return to the same or some other island. Although the number of fluctuating fragments on the surface in this case is relatively small (see Fig.~\ref{fig:fig4}a and Fig.~\ref{fig:fig4a}), the fluctuations of $\langle R_{max}\rangle$ become considerable because at these conditions small islands can be spontaneously created but most of them disappear just after several simulation time steps. Thus, for example, for $t_1=3.484$~s $\langle R_{max}\rangle_1$=21.5~nm, while for $t_2=3.485$~s $\langle R_{max}\rangle_2$=33.4~nm. The change of the maximal radius $\Delta\langle R_{max}\rangle$ in this case is 11.9~nm within 1~ms time interval. This happens because for the time frame $t_1$ there are $N_S^{(1)}=6$ single particles and $N_L^{(1)}=10$ fragments of a larger size with approximately equal diameter on the surface. For the time frame $t_2$ the number of large fragments is $N_L^{(2)}$, still equal to 10, while there are no single particles on the surface (i.e. $N_S^{(2)}=0$). With $R_L^{(1)}=R_L^{(2)}=R_L$ being the characteristic radius of the large island, $R_S^{(1)}=R_S^{(2)}=R_S$ the radius of a single particle, and $N_L^{(1)}=N_L^{(2)}=N_L$, one derives

\begin{equation}
\Delta\langle R_{max}\rangle =\frac{\Delta N_SN_L}{N_1N_2}\left(R_L-R_S\right),
\label{eq:dRmaxGeneral}
\end{equation}

\noindent
where $\Delta N_S=N_S^{(1)}-N_S^{(2)}$ is the change of the number of single particles, $N_1=N_L+N_S^{(1)}$ is the total number of particles at instance $t_1$, and $N_2=N_L+N_S^{(2)}$ is the total number of particles at instance $t_2$. Substituting values for $\Delta N_s$, $N_L$, $N_1$ and $N_2$ in Eq.~(\ref{eq:dRmaxGeneral}) for the special case considered one obtains

\begin{equation}
\Delta\langle R_{max}\rangle =\frac{15}{42}\left(R_L-R_S\right).
\label{eq:dRmax}
\end{equation}

\noindent
Substituting $\langle R_L\rangle$=32~nm and $R_S=1.25$~nm in Eq.~(\ref{eq:dRmax}), one derives $\Delta\langle R_{max}\rangle=11$~nm. Equation~(\ref{eq:dRmax}) shows that $\Delta\langle R_{max}\rangle$ increases with $R_L$ which grows with time until it reaches the equilibrium value. Equation~(\ref{eq:dRmaxGeneral}) can also be rewritten as

 \begin{equation}
\Delta\langle R_{max}\rangle =\frac{\Delta N_SN_L}{N_1^2\left(1-\Delta N_S/N_1\right)}\left(R_L-R_S\right),
\label{eq:dRmaxGeneralMod}
\end{equation}

\noindent
which shows that for $\Delta N_S\lesssim N_1$ the fluctuation of the average radius $\Delta\langle R_{max}\rangle$ can be several times larger than the the value of the average radius. Note that although the largest islands are observed for the model parameter $E_b=2$~kT (see Fig.~\ref{fig:fig3}), the largest average maximal radius is expected for $E_b=3$~kT as depicted in Fig.~\ref{fig:fig4a}. This happens because the number of single particles on the surface for $E_b=3$~kT is about 10, while for $E_b=2$~kT it is exceeding 100.

Figure~\ref{fig:fig4}e shows some dependence of $\langle R_{max}\rangle$ on $\Delta\epsilon$. The growth of $\Delta\epsilon$ leads to the decrease of $\langle R_{max}\rangle$, which is a natural result of a lower peripheral mobility of particles.

\begin{figure}
\includegraphics[width=163mm,clip]{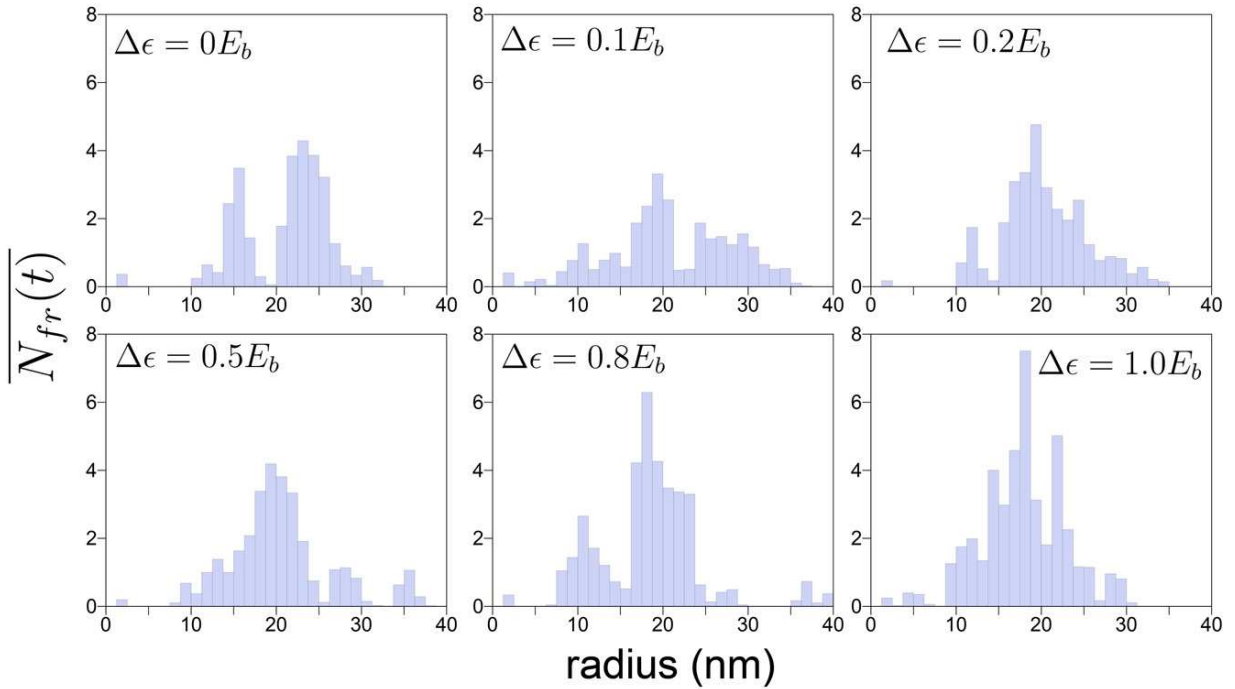}
\caption{Distributions of island sizes formed on the substrate after 4~s of simulation. The distributions were calculated at the fixed values of $E_{b} = 4$~$kT$, and $\Delta\mu = 2$~$kT$ for different values of $\Delta\epsilon$ as indicated.}
\label{fig:fig5}
\end{figure}

Figure~\ref{fig:fig5} shows the distributions of island sizes in the system after 4~s of simulation. In order to improve the statistics, the distributions shown in Fig.~\ref{fig:fig5} have been averaged over a time interval $\tau=0.78$~s as follows

\begin{equation}
\overline{N_{fr}(t)}=\frac{1}{\tau}\int_{-\tau/2}^{\tau/2}N_{fr}(t-x)dx.
\label{eq:average}
\end{equation}

\noindent
The histograms in Fig.~\ref{fig:fig5} have been calculated with different barrier energies. The maxima in the distributions show the most abundant island sizes. Figure~\ref{fig:fig5} shows that the sizes of the islands created in the fractal post-growth fragmentation process depend strongly on the binding energy $E_b$ and the barrier energy $\Delta\epsilon$. At some values of $E_b$ and $\Delta\epsilon$ one can identify two maxima in the island size distributions. Especially clear this feature manifests itself at $\Delta\epsilon=0$~$kT$, $\Delta\epsilon=0.4$~$kT$ and $\Delta\epsilon=3.2$~$kT$. The presence of two maxima in the island size distributions tells that there are two groups of islands on the surface having different preferential island size.

\begin{figure}
\includegraphics[width=163mm,clip]{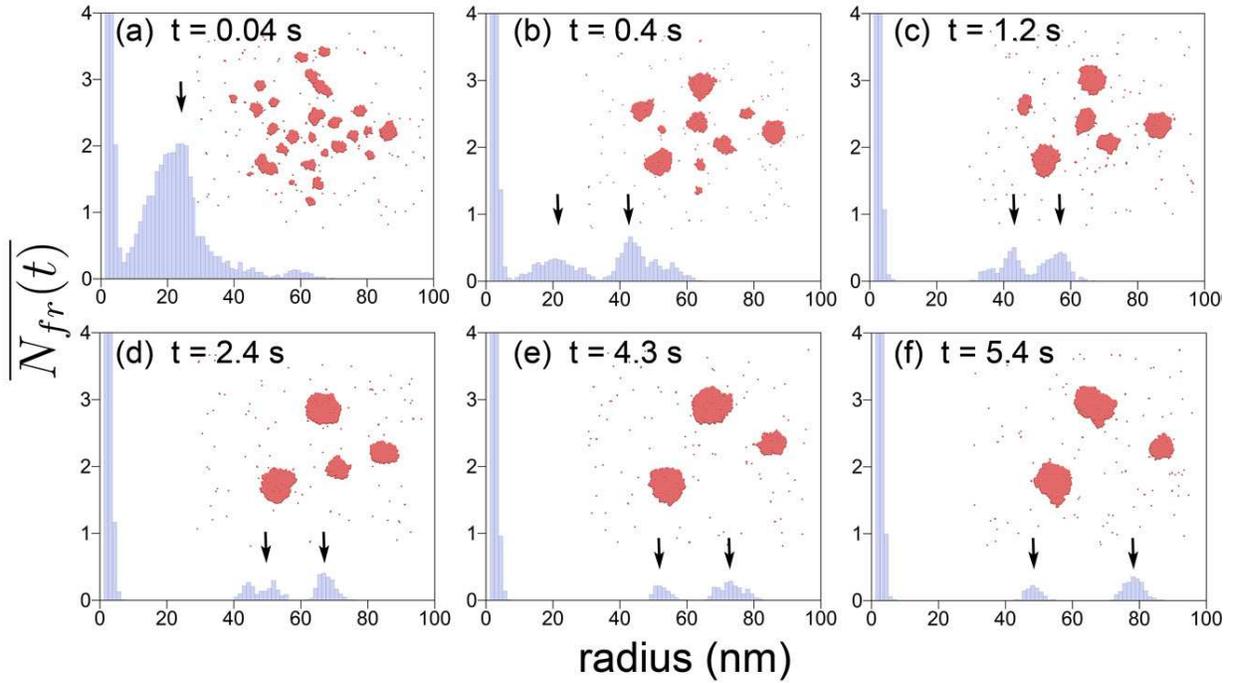}
\caption{Size distributions of islands calculated at different stages of the fractal fragmentation (see Fig.~\ref{fig:fig4}) for $E_{b} = 2~kT$, $\Delta\epsilon = 0.4~kT$ and $\Delta\mu= 2~kT$. The corresponding simulation time is given in the insets to the plots.}
\label{fig:fig6}
\end{figure}

\begin{figure}
\includegraphics[width=163mm,clip]{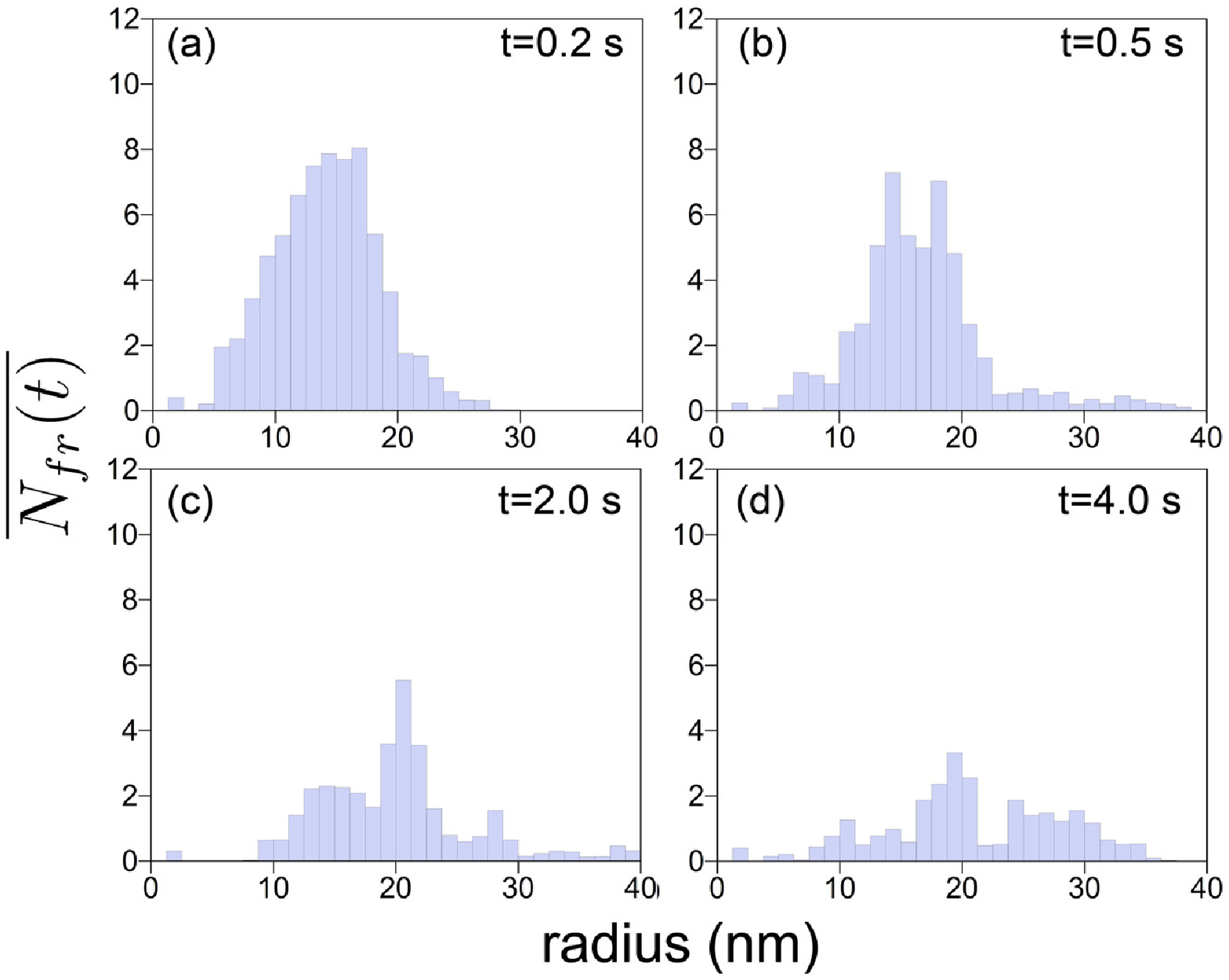}
\caption{Similar to Fig.~\ref{fig:fig6}, size distributions of islands calculated at different stages of the fractal fragmentation (see Fig.~\ref{fig:fig4}) for $E_{b} = 4~kT$, $\Delta\epsilon = 0.4~kT$ and $\Delta\mu= 2~kT$. The corresponding simulation time is given in the insets to the plots.}
\label{fig:fig7}
\end{figure}

Let us also analyze the time evolution of the distributions shown in Fig.~\ref{fig:fig5}. Figures~\ref{fig:fig6} and \ref{fig:fig7} show distributions of the island sizes calculated at different fragmentation stages for a fixed set of the model parameters. Figure~\ref{fig:fig6} illustrates the evolution of the island size distribution simulated at $E_b=2$~$kT$ and $\Delta\epsilon = 0.4~kT$. After fast fragmentation of the fractal into a subset of noncompact islands which occurs on the time scale greater than 0.04~s, the distribution of islands sizes has a Gaussian-like shape with the maximum  centered at $25$~nm. In the course of the fractal fragmentation process the magnitude and the position of the maximum of the distribution change, because the morphology of the system changes due to the evaporation of single particles from the islands and the nucleation of single particles.  Figure~\ref{fig:fig6} illustrates that small islands nucleate into larger droplets resulting in a shift of the maximum of the distribution towards larger island sizes. Interesting that the fragmentation/nucleation dynamics leads in this case of study to the formation of two maxima which correspond to the presence in the system of the droplets of different radii.

Figure~\ref{fig:fig6} shows the evolution of the fractal fragmentation process. The initial fragmentation of the fractal is very rapid. It involves the rearrangement of single particles in the fractal which form the defects at the fractal periphery. The evolution of the shape of the large droplets slows down with the growth of their size due to the decrease of the droplets mobility (see Figs.~\ref{fig:fig6}b-c). At the stage when only a few large-size droplets remain their dynamics is governed to large extend by the interchange of peripheral particles from these droplets (see Figs.~\ref{fig:fig6}d-f). The large droplets diffuse slowly over a surface and may eventually merge. The characteristic time scale for diffusion of an entire large droplet is significantly larger than the characteristic diffusion time of single constituent particle, and therefore practically can not be resolved within the simulation time limit. However, note that this motion can also be simulated with a larger time step. The appropriate value of the time step can be estimated using Eq.~(\ref{eq:eq5}).

Figure~\ref{fig:fig7} shows the slower evolution of the island size distribution as compared to Fig.~\ref{fig:fig6}. Slowing down of the process is caused by the increase of the binding energy $E_b$ between particles within the fractal. Figure~\ref{fig:fig7} shows that as in Fig.~\ref{fig:fig6} a Gaussian-like distribution of a large number of droplets arises immediately after the fractal fast fragmentation. The maximum of this distribution slowly drifts towards the larger droplet sizes as the smaller islands nucleate (see Fig.~\ref{fig:fig7}b). Remarkably, that at the later stages ($t=4$~s, Fig.~\ref{fig:fig7}d) several maxima arise in the distribution. It is worth noting that this feature of the droplet size distribution was also observed in experiment \cite{BrechignacDyson10}.

Another useful quantity for the characterization of surface structures is the ratio between the area and the perimeter of the structure ($S/P$ ratio) \cite{Brechignac06}. This ratio characterizes the island topology. Thus, the $S/P$ ratio for a linear chain of $N$ spherical particles is equal to

\begin{equation}
\frac{S}{P} = \frac{d_{0}}{4},
\label{eq:eq18}
\end{equation}

\noindent
where $d_{0}$ is the diameter of a particle. Note that the $S/P$ ratio for a linear chain is always a constant.
The $S/P$ ratio for a compact droplet of the radius, $R_{d}$, is equal to

\begin{equation}
\frac{S}{P} = \frac{R_{d}}{2}.
\label{eq:eq20}
\end{equation}

\noindent
It can be easily expressed via the number of particles $N$ in the droplet:

\begin{equation}
\frac{S}{P} = \frac{d_{0}}{4}\sqrt{N}.
\label{eq:eq19}
\end{equation}

\noindent
In this case the $S/P$ ratio increases as $\sqrt{N}$ with the growth of the system size. The $S/P$ ratio for a fractal consisting of $N$ particles should be larger than in Eq.~(\ref{eq:eq18}) and smaller than in Eq.~(\ref{eq:eq19}). Let us now analyze the time evolution of the average $\langle S/P\rangle$ ratio of the system during the fractal fragmentation. The $\langle S/P\rangle$ ratio for a system of $N$ islands is defined as

\begin{equation}
\langle S/P \rangle = \frac{1}{N_{fr}} \sum^{N_{fr}}_{i=1} \frac{S_{i}}{P_{i}},
\label{eq:eq21}
\end{equation}

\noindent
where $S_{i}$ and $P_{i}$ is the area and the perimeter of $i$-th island, and $N_{fr}$ is the number of islands in the system. The $\langle S/P\rangle$ ratio is a useful characteristic for the structures morphology, often used in experiment \cite{Brechignac06}.

The dependence of the $\langle S/P\rangle$ ratio on time calculated for different sets of the model parameters is shown in Fig.~\ref{fig:fig4}c and Fig.~\ref{fig:fig4}f. Curve 1 in Fig.~\ref{fig:fig4}c shows time evolution of the $\langle S/P\rangle$ ratio during the fractal relaxation in the case of the relatively small binding energy between the particles being equal to $1~kT$. The $\langle S/P\rangle$ ratio in this case rapidly decreases until it reaches the minimum value $0.78$~nm, i.e. the $S/P$ ratio which is slightly smaller than the value for a dimer of particles with $d_{0} =2.5$~nm. Figure~\ref{fig:fig4} shows that the $\langle S/P\rangle$ dependencies to large extend follow the dependencies calculated for $\langle R_{max}\rangle$.

\begin{figure}[tb]
\includegraphics[width=163mm,clip]{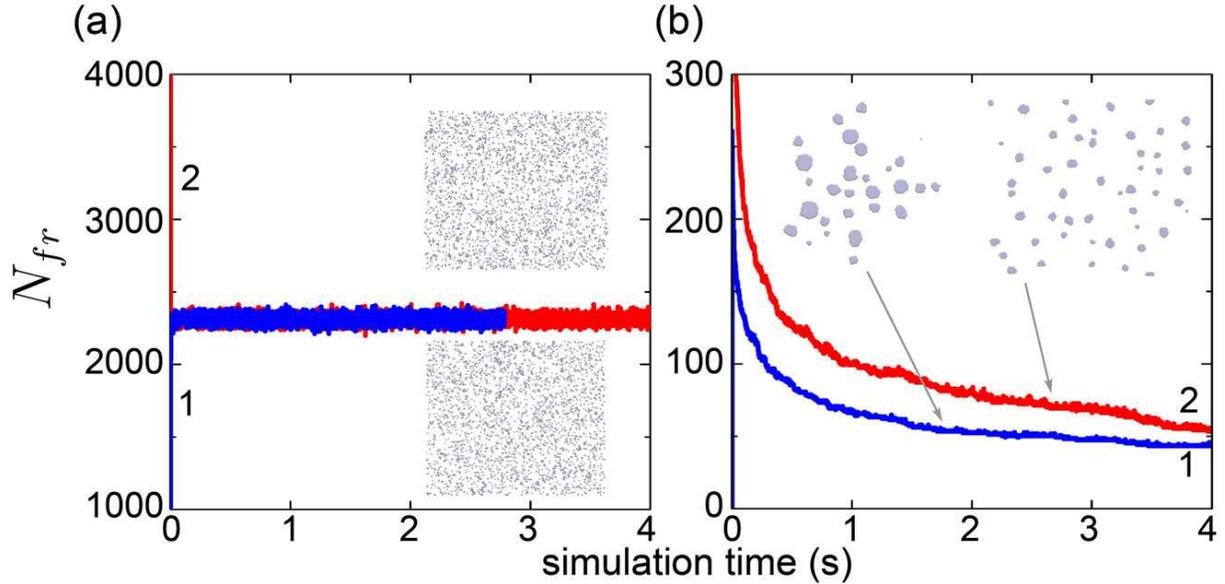}
\caption {Time evolution of the number of fragments/nucleation islands on a surface, $N_{fr}$, during the fractal fragmentation process (line 1) and during the nucleation process of randomly distributed particles (line 2). The calculations have been performed for a $650\times750$~nm$^2$ substrate with periodic boundary conditions. Plots (a) and (b) have been calculated at different values of the model parameters: (a) $E_{b}=1$~$kT$, $\Delta \epsilon = 0.2$~$kT$, $\Delta\mu = 2$~$kT$; (b) $E_{b}=4$~$kT$, $\Delta \epsilon = 0.4$~$kT$, $\Delta\mu = 2$~$kT$.  The insets show the morphology of the system at the end of the simulation.}
\label{fig:equilibriumNfr}	
\end{figure}

The performed analysis provides a lot of useful information on the dynamical evolution of the system during fragmentation. However, its direct comparison with experimental measurements is rather difficult because the calculated distributions vary with time, but the experimental measurements are usually performed for stationary (or quasi-stationary) systems. Nevertheless the comparison with experiment is possible if the average life-time $T_l$ of the studied configuration is greater than the characteristic measurement time $T_m$:

\begin{equation}
T_{l}\gtrsim T_{m}.
\label{eq:measureCond}
\end{equation}

\noindent
Here, $T_l$ is defined as the characteristic time-period at which an observable characteristic, e.g., the number of fragments in the system, changes within the statistical uncertainty, and $T_m$ is the minimal time-period required to perform an experimental measurement.

An important characteristic of the system's stability, is the total number of fragments $N_{fr}$ in the system. At the equilibrium $N_{fr}$ fluctuates around the average constant value. Note, that $N_{fr}$ may have similar behavior in a so-called kinetically trapped state, or a quasi-equilibrium state which is separated from the equilibrium state by an energy barrier. The energy barrier between the kinetically trapped state and the equilibrium state may be significantly larger than the thermal vibration energy, therefore the trapped system may spend a noticeable lifetime in the kinetically trapped state. This life-time can be sufficient for experimental measurements and for holding Eq.~(\ref{eq:measureCond}). This means that the quasi-equilibrium value of $N_{fr}$ may come out different for different initial distributions of particles on a surface, demonstrating that different evolution paths may lead the system to different final quasi-equilibrium states. Below we analyze two examples supporting this hypothesis.

\begin{figure}[tb]
\includegraphics[width=163mm,clip]{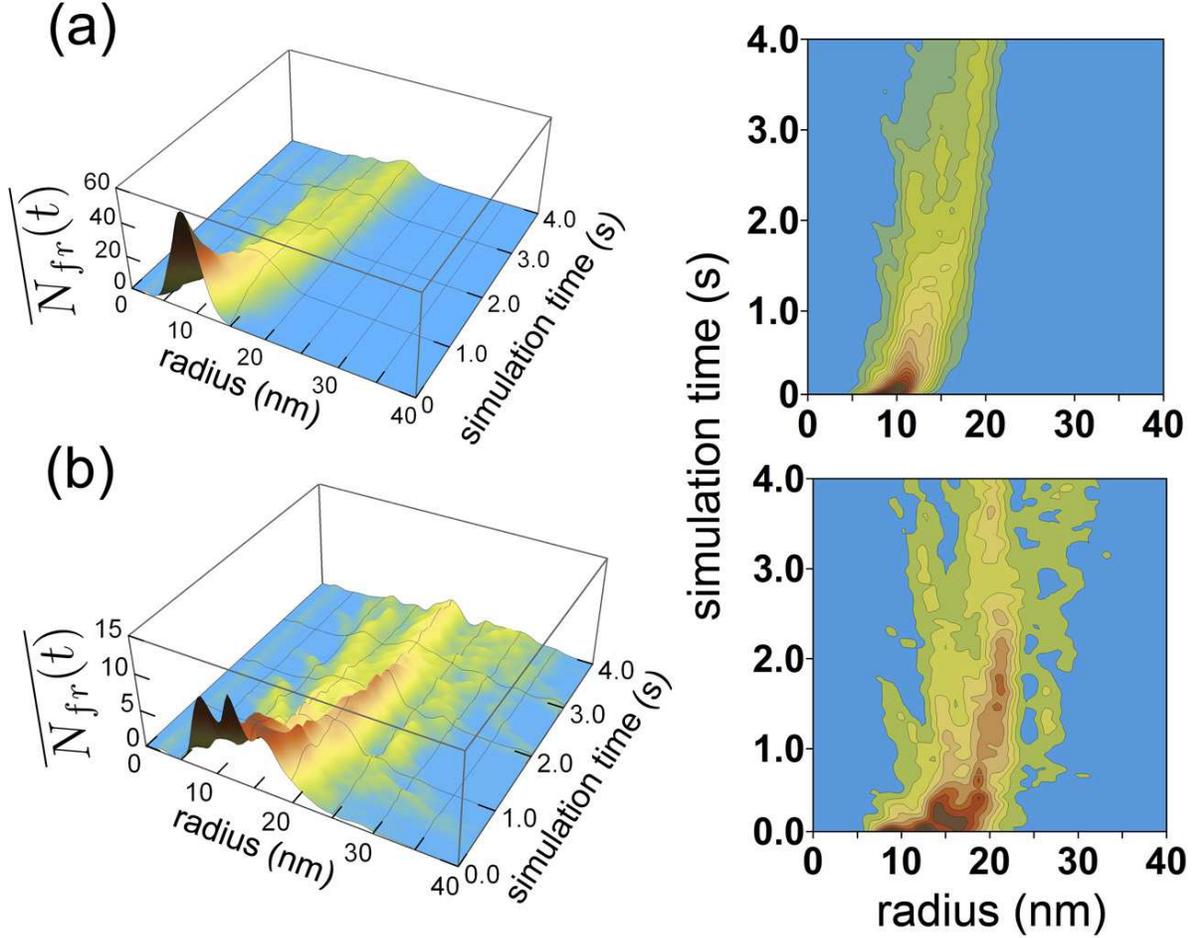}
\caption{Time evolution of the island size distributions calculated for the nucleation process of randomly distributed particles (plot a) and for the fractal fragmentation process (plot b). The distributions have been calculated for the same values of the model parameters as in Fig.~\ref{fig:equilibriumNfr}b. The initial fractal shape has been chosen the same as in Fig.~\ref{fig:fig2}a.}
\label{fig:EquilDistr}	
\end{figure}

Figure~\ref{fig:equilibriumNfr} depicts the time evolution of the number of fragments/nucleation islands, $N_{fr}$, calculated (line 1) for the fractal having the initial shape as plotted in Fig.~\ref{fig:fig2}a, and (line 2) during the nucleation process of randomly distributed particles. The total number of constituent particles in both cases is equal to 5182. The size of the substrate used in the simulation is identical in both cases, equal to $650\times750$~nm$^2$. Figure~\ref{fig:equilibriumNfr} shows that the inter-particle interaction influences significantly the system dynamics. Thus, in the case of the weak bonding between particles (i.e. $E_{b}=1$~$kT$, $\Delta \epsilon = 0.2$~$kT$), see Fig.~\ref{fig:equilibriumNfr}a line 1, the fractal fragments into $\sim$2320 islands, i.e. most of the particles in the system are bound in a form of dimers. Remarkably, that at these model parameters particles randomly distributed over a surface nucleate to approximately the same quasi-equilibrium value $N_{fr}$ (line 2 in Fig.~\ref{fig:equilibriumNfr}a). The insets in Fig.~\ref{fig:equilibriumNfr}a illustrate the distribution of particles at the instant $t=4$~s in the case of nucleation and at $t=2.8$~s for the fragmentation. Figure~\ref{fig:equilibriumNfr}a shows that the system can evolve from the very different initial states to the same final state.

The fragments number evolution with time depend on the inter-particle interaction as seen from Fig.~\ref{fig:equilibriumNfr}b, obtained at larger $E_b$, $E_{b}=4$~$kT$, $\Delta \epsilon = 0.4$~$kT$. The quasi-equilibrium value of $N_{fr}$ in this case depends on the initial distribution of particles on a surface. The inset to Fig.~\ref{fig:equilibriumNfr}b shows that both systems have evolved in a group of droplets, whereby the size of the droplets created from the initial fractal distribution of particles is larger than the size of the droplets created via the nucleation.

Figure~\ref{fig:equilibriumNfr} shows that for the chosen model parameters the number of fragments in the system becomes constant or changes slowly with time at sufficiently large $t$ value. The resulting static or quasi-static distributions of particles can be compared with experimental observations. In the cases when the initial distribution of particles on a surface influences the final morphology of the system means the system occupies one of the kinetically trapped state. Although the quasi-equilibrium kinetically trapped states do not have the lowest free energy, they may live for sufficiently long time to perform experimental measurements of the system characteristics. The asymptotic behavior of the fragments distribution with time is well seen in Fig.~\ref{fig:EquilDistr}. Figure~\ref{fig:EquilDistr} shows the time evolution of the island size distributions calculated for the processes depicted in Fig.~\ref{fig:equilibriumNfr}b. The island size distribution characterizing the period 0-2~s experiences significant variation, while the distribution during 2-4~s is almost static, with only a minor change.

\begin{figure}
\includegraphics[width=163mm,clip]{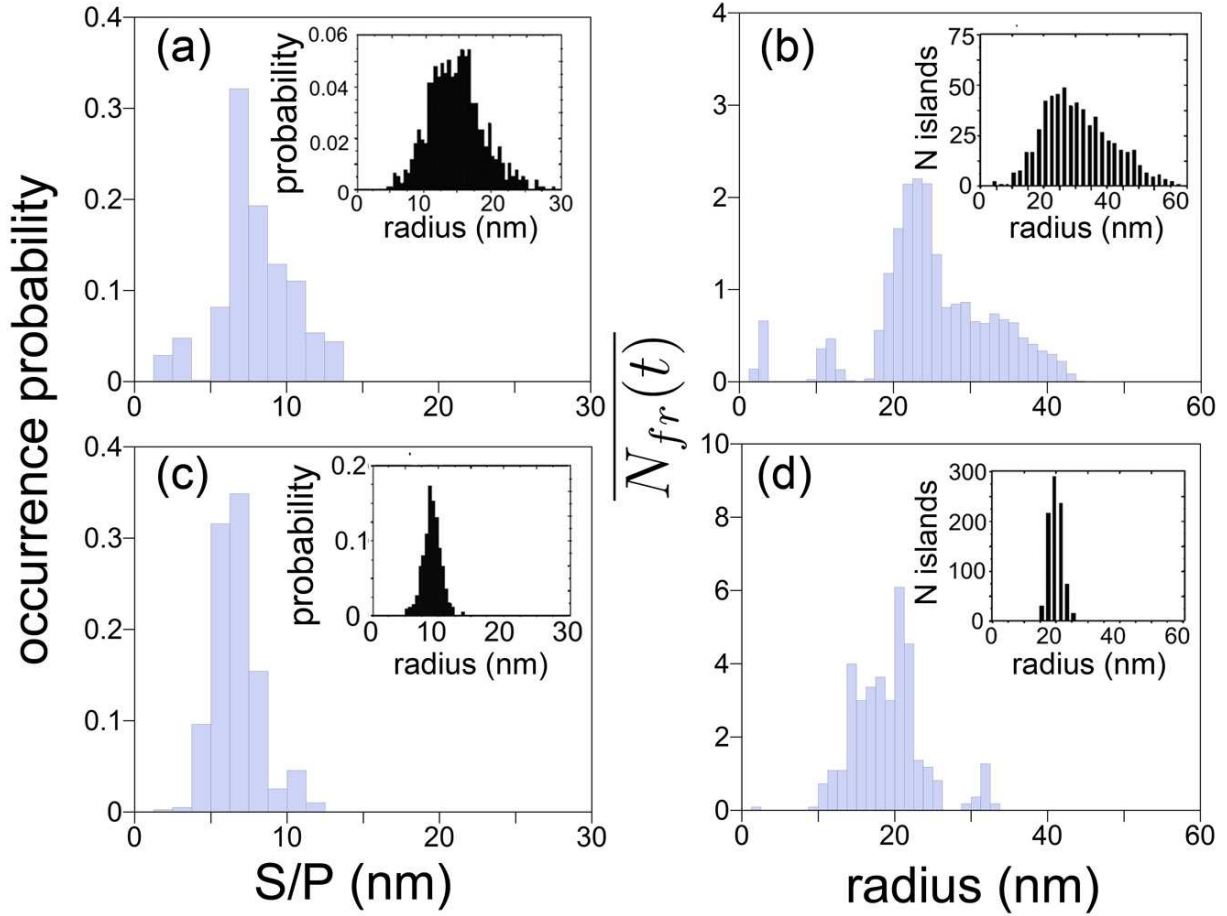}
\caption{$S/P$ ratio distributions calculated after the fractal fragmentation with different sets of the model parameters and the corresponding distributions of island sizes. Distributions $(a)$ and $(b)$ are calculated with $E_{b}=3$~$kT$, $\Delta\epsilon = 0.6~E_{b}$, $\Delta\mu = 10$~$kT$; $(c)$ and $(d)$ with $E_{b}= 4$~$kT$, $\Delta\epsilon = 0.4~E_{b}$, $\Delta\mu = 2$~$kT$. Insets show the results of experimental measurements for silver fractal fragments created via annealing $(a)$ and $(b)$, and by adding of oxide impurities to silver clusters $(c)$ and $(d)$~\cite{Brechignac06}.}
\label{fig:fig8}
\end{figure}

Figure~\ref{fig:fig8} shows the island size distributions and the corresponding $S/P$ ratio distributions calculated for the fractal fragmentation on the $650\times750$~nm$^2$ with periodic boundary conditions. The distributions plotted in Figs.~\ref{fig:fig8}a-b have been obtained with the model parameters $E_{b}=3$~$kT$, $\Delta\epsilon = 0.6~kT$, $\Delta\mu = 10$~$kT$ at $t=4$~s, i.e. well after the fractal fragmentation when the system evolves in the almost stationary equilibrium or quasi-equilibrium state. In this case diffusion of particles along the fractal periphery is the dominating process. The increased rate of particle peripheral diffusion leads to the faster island rearrangement, and the formation of islands of different size, as seen in Fig.~\ref{fig:fig8}b. The insets in Fig.~\ref{fig:fig8}a-b show the results of experimental measurements obtained for silver fractals fragmentation via annealing at $600$~K. The experimentally measured distribution of the silver cluster island sizes is rather broad, with the most probable radius of silver islands $\sim25$~nm. A close value of 23~nm follows from the theoretical analysis. The discrepancy may arise due to the thinner branches of the fractal used in the simulations as compared to the ones analyzed in experiment.

Figures~\ref{fig:fig8}c-d show the island size distribution and the corresponding $S/P$ ratios distributions calculated with $E_{b}=4~kT$, $\Delta \epsilon=0.4~kT$, $\Delta \mu=2~kT$. The results of numerical calculation are compared with the experimental data shown in the insets to Figs.~\ref{fig:fig8}c-d on silver fractals grown with the oxidized silver nanoparticles~\cite{Brechignac06}. In the experiment the most abundant radius of the silver cluster islands is $18$~nm, being in good agreement with the results of our calculations as seen from Fig.~\ref{fig:fig8}d.

Note that the width and the position of the maximum in the calculated distributions shown in Fig.~\ref{fig:fig8} are rather close to the experimentally observed ones while the absolute value of the experimental and theoretical distributions differ quite significantly. This happens because we analyze the dynamics of a single fractal, while the experimental measurements deal with many fractals on a surface.

\section{Conclusion}

We have performed theoretical analysis of the post-growth processes occurring in a nanofractal on a surface using the method, which models the internal dynamics of particles in a fractal and accounts for their diffusion and detachment. We have demonstrated that these kinetic processes are responsible for the shape of the islands created on a surface after the post-growth relaxation.

The suggested theory is general and can be used in studies of the formation and relaxation processes of different nanostructures deposited on a surface. The developed model includes three parameters, which are determined by interatomic interactions in the system and could in principle be theoretically calculated for each particular case on the basis of the full atomistic approach for the dynamics of a single particle on a surface. The model parameters can also be obtained from experiment and are specific for different types of substrates and deposited materials. In the present paper we analyze the fractal dynamics on a surface at various values of the model parameters within a wide range of values and reveal the main fragmentation scenarios of the system.

The paper presents a significant advance in the understanding of paths of the fragmentation of deposited nanosystems. It opens a broad spectrum of questions for further investigations. Thus, it is interesting to explore the link of the model parameters with the structural properties (both electronic and geometrical) of the deposited particles and substrates as well as their thermal, mechanical, electromagnetic, etc. properties. Thus, for instance introduced model parameters can be determined from the molecular dynamics simulations of different diffusion processes occurring on a surface.

In the performed analysis the deposited particles are assumed to be stiff, i.e. without any internal degrees of freedom. However, the particle diffusion over a surface may change quite significantly when the particle experiences deformation or changes its phase state. Accounting for the detail internal structure and dynamics of particles in the context of their diffusion is one of the next obvious steps towards the better understanding of the very complex process discussed in this work.

In the present paper we have studied particle dynamics in 2D. Another obvious important extension of the model is to investigate the role of the third dimension in the process of fractal formation and fragmentation. This is especially interesting to do, because there are many examples of three dimensional fractal systems in biology \cite{Mashiah08, Baish00}, where the dendritic shapes are rather common. Understanding of the growth evolution and fragmentation of such systems is very important and may have applications in medicine

\begin{acknowledgments}
This work was supported by the European EXCELL project. The possibility to perform complex computer simulations at the Frankfurt Center for Scientific Computing is also gratefully acknowledged.
\end{acknowledgments}

\bibliographystyle{apsrev}
\bibliography{my_bib}

\end{document}